\documentclass[aps,superscriptaddress,twocolumn,showpacs,preprintnumbers,amsmath,amssymb,showkeys,textcomp]{revtex4}
\usepackage[cp1251]{inputenc}
\usepackage[english]{babel}
\usepackage{textcomp,amssymb,amsmath}
\usepackage{amstext,textcomp} 
\usepackage[dvips]{hyperref}
\usepackage[mathscr]{eucal}
\usepackage{longtable}
\usepackage{graphicx}
\usepackage[english]{babel}

\begin{document}
\title{Ferromagnetic  Ordering in Carbon Nanotubes, Incorporated in Diamond Single Crystals}

\author{Dmitri Yerchuck (a),  Vyacheslav Stelmakh (b), Alla Dovlatova (c), Yauhen Yerchak (b), Andrey Alexandrov (c)\\
(a) - Heat-Mass Transfer Institute of National Academy of Sciences of RB, Brovka Str., 15,  Minsk, 220072,  RB,\\ dpy@tut.by\\
(b) Belarusian State University, Nezavisimosti Avenue 4, Minsk, 220030, RB
(c) - M.V.Lomonosov Moscow State University, Moscow, 119899}
\date{\today}
             
\begin{abstract}  The physical origin of the mechanism of the formation of ferromagnetic ordering in carbon nanotubes (NTs), produced by high energy ion beam modification of diamond single crystals in $\langle{110}\rangle$ and $\langle{111}\rangle$ directions has been found. It is concluded from analysis of experimental results on ferromagnetic spin wave resonance observed, that the only  $\pi$-electronic subsystem of given NTs is responsible for the appearance of ferromagnetism. It is determined by asymmetry in spin density distribution in Su-Schrieffer-Heeger (SSH) topological soliton lattice. The formation of SSH topological soliton lattice is considered in the frames of generalized SSH-model of organic conductors, in which  $\pi$-electronic subsystem is represented being to be 1D quantum Fermi  liquid.
 
\end{abstract}

                             
\maketitle
\section{Introduction and Background}
It is well known that all substances on the whole are magnetics. At the same time, it is also well known, that classical magnetic ordering is existing in the substances, which are built from the  atoms with unfilled inner atomic $d$- or $f$-shells or include given atoms in their elementary units. In other words, classical  magnetics are the substances,  elementary units of which include transition chemical elements with unfilled atomic 3d-, 4d-, 5d-, 6d-shells, or rare earth elements with unfilled atomic 4f, 5f-shells.  Carbon does not refer to given group. Nevertheless, there are at present a number of  communications on  magnetic ordering in carbon and carbon based materials. 

The first report on the experimental revealing of magnetic ordering in carbon structurally ordered systems  was presented on the IBMM-Conference in Knoxville, TN, USA \cite{Erchak_Knoxville} in 1990.  Given result was confirmed in report on E-MRS Conference in Strasbourg, France \cite{Efimov} also in 1990. Let us remark, that   the first report on magnetic ordering in structurally non-ordered carbon materials is appeared almost in the same time. It is  the work \cite{Kawataba}, where ferromagnetic ordering in pyrolytic carbon, produced  by chemical vapour deposition
(CVD) method using adamantane to be source  material, was found.  Let us also remark, that simultaneously, the reports \cite{Erchak_Knoxville}, \cite{Efimov} were the first reports on the formation by high energy ion beam modification (HEIBM) of diamond single crystals  structurally and magnetically ordered quasi-one-dimensional (quasi-1D) system along ion tracks, that is new carbon allotropic form, which was identified with nanotubes {NTs}, incorporated in diamond matrix in direction, precisely coinciding with ion beam direction.  Given system possesses by a number of interesting physical properties, reported in \cite{Erchak}, \cite{Erchak_JETP}, \cite{Ertchak_JAS},
 \cite{Ertchak_Stelmakh}. When concerne the magnetic ordering, it was established from the study of temperature dependence of electron spin resonance (ESR) absorption intensity, that, for instance, incorporated nanotubes, produced by neon HEIBM of diamond single crystal along  $\langle{100}\rangle$ crystallographic direction, possess by weak antiferromagnetic ordering \cite{Erchak}, \cite{Ertchak_JAS}, \cite{Ertchak_Stelmakh}. At the same time, copper HEIBM with implantation direction along  $\langle{111}\rangle$ crystal axis,  nickel HEIBM with implantation direction along $\langle{110}\rangle$  axis, \cite{Erchak}, \cite{Ertchak_JAS}, \cite{Ertchak_Stelmakh},  and  boron HEIBM of polycrystalline diamond films with implantation direction transversely  to film surface, \cite{Erchak_JETP}, lead to formation of NTs, incorporated in diamond matrix, which are possessed by ferromagnetic ordering. It was established directly by observation of ferromagnetic spin wave resonance (FMSWR) \cite{Erchak_JETP}, \cite{Ertchak_JAS}, \cite{Ertchak_Stelmakh}. The first observations of FMSWR in carbon materials and in the materials, which are non-traditional magnetic materials, on the whole, were reported in above cited works. It was established, that magnetic ordering is inherent property for carbon electronic system, that is,  it is not connected with presence of magnetic impurities, since starting samples were selected in that way, that the absolute spin number of  paramagnetic impurities and paramagnetic structural imperfections at all did not exceed the value ~ $10^{12}$  spins. 

Very recently, \cite{Yerchuck_D_Dovlatova_A}, antiferroelectric ordering has been found in the same pure carbon allotropic form - quasi-1D carbon zigzag-shaped  nanotubes (CZSNTs), obtained by boron- and copper-HEIBM of diamond single crystals in $\langle{111}\rangle$-direction. It was established  by means of the detection of new optical phenomenon - antiferroelectric spin wave resonance (AFESWR), which was theoretically described and experinmentally confirmed for the first time by infrared (IR) spectroscopy study of carbynes  in  \cite{Yearchuck_PL}. It seems to be very new property of pure carbon allotropic forms  - quasi-1D CZSNTs  and  carbynes.  Moreover, on the observation of  antiferroelectric  ordering has been reported in \cite{Yearchuck_PL} for the first time  for all  carbon and carbon based systems on the whole. Given results mean, that pure carbon in the form of
quasi-1D CZSNTs  or carbynes  is multiferroic system. Especially significant was the observation of  AFESWR with linear $k$-dipersion law, where  $k$ is magnitude of wave vector $\vec{k}$. It was the first work, in which the general spin wave theory was experimentally confirmed. Let us remark, that given  theory was starting from the work \cite{Hulthen}  in 1936 and was develoved in many subsequent works, for instance, in \cite{Anderson},  \cite{F_Keffer},  \cite{Keffer}, \cite{Ziman}, \cite{Nakamura}, \cite{Tani}, however for antiferromagnetic spin wave case. At the same time,  it has been shown in \cite{Yearchuck_PL}, that the conclusions of the  antiferroelectric spin wave  theory and antiferromagnetic spin wave theory are qualitatively identical, in particular, dipersion law is the same.  

Let us also remark, that experimental observation of multiferroicity in quasi-1D CZSNTs and  carbynes  means the breakdown of space inversion symmetry  along  CZSNT hypercomplex  symmetry axis  and respectively along  carbyne chain  symmetry axis. In the case of  CZSNTs, it agrees well with the model of quasi-1D CZSNTs \cite{Dovlatova_Yearchuck}, \cite{Yerchuck_D_Dovlatova_A}, based on bond dimerization in  all chain components of  quasi-1D CZSNT along its  hypercomplex symmetry axis $z$, which actually leads to inversion symmetry breakdown along given axis. Therefore, the experimental observation of antiferroelecricity of quasi-1D CZSNTs, necessary condition for which is the evident prediction of the model (appearance of nonzero polarisation by atomic displacements), proposed in \cite{Dovlatova_Yearchuck}, can be considered to be additional argument in favour of given  theoretical model. 
 The idea of the formation of ferromagnetic ordering (or ferroelectric ordering) in quasi-1D carbon systems was proposed in \cite{Yearchuck_PL}. It has been shown, that by asymmetric deviation of spin  density relatively
the chain direction the constant component is appeared. It
can lead in the case of density wave formation along 1D axis to ferromagnetic ordering, if density wave
distribution is magnetic spin distribution and to ferroelectric ordering
by electric own moments or electric dipole distribution. At the same time, which subsystem (or subsystems), that is $\pi$ subsystem or $\sigma$ subsystem (or even both subsystems) is (are) responsible for ferromagnetic ordering in quasi-1D CZSNT  was not established.

 The aim of given work is to study in more details the properties of cylindrical nanotubes, produced in diamond single crystals by high energy ion implantation, which are  possessing  by $C_\infty$ symmetry axis,  and to  experimentally establish the mechanisms of formation of ferromagnetic and ferroelectric ordering in given NTs.

 \section{Experimental Technique}
Samples of type IIa natural diamond, implanted by high energy ions of nickel (the energy of ions in ion beam was  $335$ $MeV$, ion beam dose was $5\times{10^{14}}$ $cm^{-2}$) and  implanted by
high energy ions of copper and boron (the energy of
ions in ion beam was 63 MeV and 13.6 MeV for copper and boron ions correspondingly, ion beam dose was
$5\times 10^{14} cm^{-2}$) have been studied. Nickel implanted sample was 
representing rectangle  with the sides $\approx$ 5 and  4 mm in based (110) plane. Copper and boron  implanted samples were 
representing in their geometry prismes with quilateral triangle in their bases, coinciding with (111) crystallographic
plane, side of base triangle was equal to $\approx$ 5mm. Thickness
of all samples was equal to $\approx$ 1mm. Ion implantation of copper and boron
was performed along [111] crystal direction, that is transversely to triangle prism base uniformly along all the surface. Nickel implantation was performed along [110] crystal direction, that is transversely to based (110) rectangle plane also  uniformly along all the  plane surface. The samples studied were magnetically pure samples, since they have been selected so that the absolute spin number did not exceed the value $\approx 10^{12}$ spins in each of the sample used  before implantation.  The temperature of the samples during the implantation was controlled and it did not exceed 400 K.

 ESR spectra were registered on X-band ESR-spectrometer "Radiopan" at room temperature by using of $TE_{102}$ mode rectangular cavity. The ruby standard sample was permanently placed in the cavity on its sidewall. One of the  lines of ESR absorption by  $Cr^{3+}$ point paramagnetic centers (PC)   was used  for the correct relative intensity measurements of ESR absorption, for the calibration  of the amplitude value of magnetic component of the microwave field and for precise phase tuning of microwave field. It was possible owing to unsaturating behavior of ESR absorption in ruby in the range of  the microwave power applied, which was  $\approx  100$ mW in the absence of attenuation. Unsaturable character of the absorption in a ruby standard was confirmed by means of the measurements of the absorption intensities in two identical ruby samples in dependence on the microwave power level. The first sample was standard sample,  permanently placed in the cavity, the second sample was placed in the cavity so that the resonance line intensity was about 0.1 
of the intensity of corresponding line of the first sample. They were registered  simultaneously but their absorption lines  were not overlapped owing to slightly different sample orientations. The  foregoing intensity ratio  was preserved for all microwave power values in the range used, which indicates, that really ruby samples are good standard samples in ESR spectroscopy studies. 

\section{Results}

There has been established, that copper and boron  HEIBM with implantation direction along  $\langle{111}\rangle$ crystal axis,    and  boron HEIBM of polycrystalline diamond films with implantation direction transversely  to film surface  lead to formation of NTs, incorporated in diamond matrix, which  possess by relatively weak ferromagnetic ordering. The experimental results were in details represented in \cite{Erchak}, \cite{Erchak_JETP}, \cite{Ertchak_JAS},
 \cite{Ertchak_Stelmakh}, and  they will be not reproduced in presented paper.
 The studies of  FMSWR in the sample implanted along  $\left\langle{110}\right\rangle$ crystal direction were  also described in \cite{Ertchak_Stelmakh}, however we will summarise in the paper presented the  experimental results with more details and some new results will be given, which seem to be essential to establish the detailed processes, leading to ferromagnetic ordering formation, since the concrete subsystem, which is responsible for the mechanism of the ferromagnetically ordered state was not proposed in the papers above cited. 

 The strong anisotropy of FMSWR excitation has been observed in implanted unannealed samples.  FMSWR was registered the only in the case, when the implantation plane was perpendicular to the vector $\vec{H}_1$ of magnetic component of the microwave field. The intensity of  FMSWR absorption was increasing with temperature of isochronal 20-minute's annealing  up to $\approx{700}$ Celsius degrees. It is interesting, that after the annealing at 350 Celsius degrees and by higher temperatures FMSWR absorption was detected also by the vector $\vec{H}_1$ of magnetic component of the microwave field position, being to be parallel to sample implantation plane. The parameters of FMSWR absorption, that is splitting parameter, which can be characterised by, for instance, effective g-values of its modes (since central FMSWR-line, which corresponds to ferromagnetic resonance has the only weak anisotropy and it can be considered approximately isotropic if to compare its anisotropy with rather strong anisotropy of FMSWR modes). We will use the  g-values of the third mode, Figure 1 for given characteristics. It is seen, that amplitudes, linewidths and intensities  of FMSWR modes are also strongly anisotropic, see Figures 2 to 4,  where angular dependencies  of the linewidth, FMSWR absorption  amplitude and intensity of the third  FMSWR-mode are presented.
\begin{figure}
\includegraphics[width=0.5\textwidth]{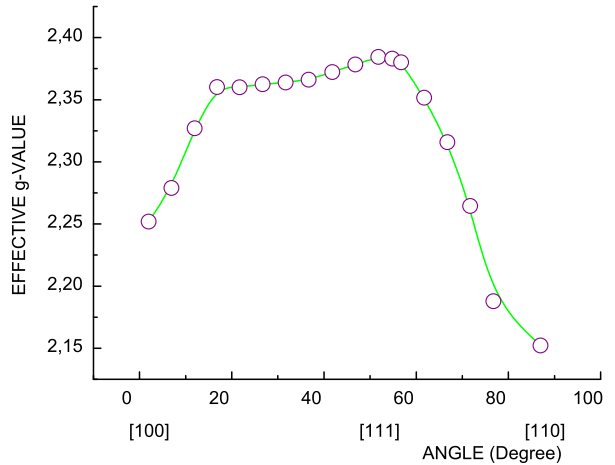}
\caption[Angular dependence of effective g-value for the third  FMSWR-mode, observed in NTs, incorporated in diamond single crystal by  nickel ion beam direction transversely (110) sample plane] {\label{Figure1}Angular dependence of effective g-value for the third  FMSWR-mode, observed in NTs, incorporated in diamond single crystal by  nickel ion beam direction transversely (110) sample plane}
\end{figure}
\begin{figure}
\includegraphics[width=0.51\textwidth]{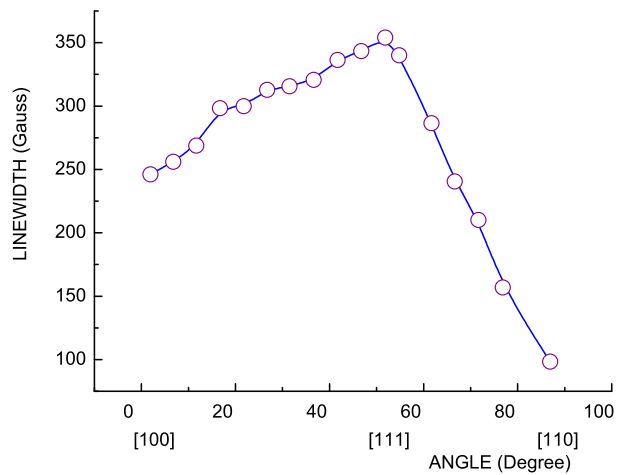}
\caption[Angular dependence  of the linewidth of the third  FMSWR-mode, observed in NTs, incorporated in diamond single crystal by  nickel ion beam direction transversely (110) sample plane ]{\label{Figure2} Angular dependence  of the linewidth  of the third  FMSWR-mode, observed in NTs, incorporated in diamond single crystal by  nickel ion beam direction transversely (110) sample plane}
\end{figure}
\begin{figure}
\includegraphics[width=0.5\textwidth]{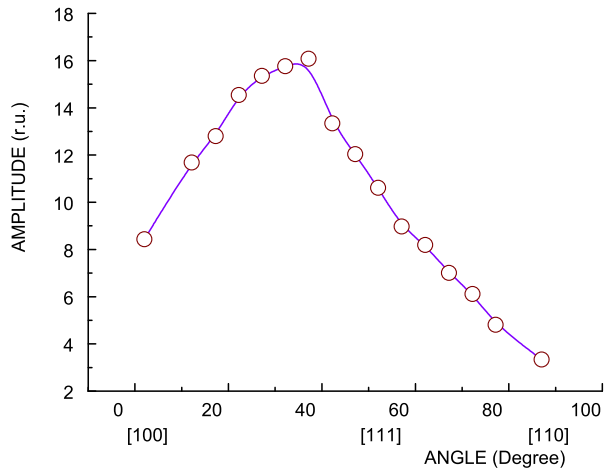}
\caption[Angular dependence of FMSWR absorption amplitude of the  the third  FMSWR-mode, observed in NTs, incorporated in diamond single crystal by  nickel ion beam direction transversely (110) sample plane ] {\label{Figure3} Angular dependence of FMSWR absorption amplitude of the  the third  FMSWR-mode, observed in NTs, incorporated in diamond single crystal by  nickel ion beam direction transversely (110) sample plane}
\end{figure}

\begin{figure}
\includegraphics[width=0.47\textwidth]{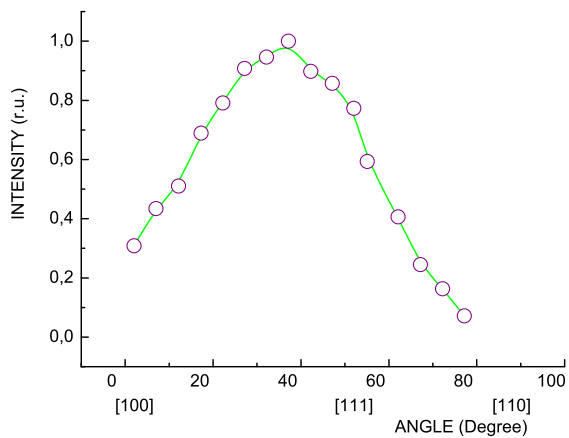}
\caption[Angular dependence  of the FMSWR absorption intensity of the third  FMSWR-mode, observed in NTs, incorporated in diamond single crystal by  nickel ion beam direction transversely (110) sample plane ]{\label{Figure4} Angular dependence  of the FMSWR absorption intensity of the third  FMSWR-mode, observed in NTs, incorporated in diamond single crystal by  nickel ion beam direction transversely (110) sample plane}
\end{figure}
The splitting parameter anisotropy seems to be nontrivial, it characterises by two maxima - at 54.7 degrees, that is, in [111] direction of diamond lattice, and in direction about 17 degrees from [100] crystal direction.  It is interesting, that  the angular dependence of linewidth has the resemblance with the angular dependence of g-value, it characterises also by two  maxima at the same angles, see Figure 2. Minimal  effective g-value  and minimal value of linewidth for the third mode are achieved in the case of $\left\langle{110}\right\rangle$ ion beam modification at [110] direction. They are equal to 2.1526 and equal to 100 G respectively, maximal effective g-value and maximal  value of linewidth correspond to [111] direction of diamond lattice and are equal respectively to 2.3845 and 360 G. Coincidence of the directions of anisotropy maxima with diamond lattice characteristic directions is indication of the role of host diamond lattice on the characteristics of magnetic ordering, determining the magnetic symmetry, that seems to be possible by relative softness of the corresponding electronic subsystems. Therefore we obtain the strong evidence, that ferromagnetic ordering is determined by electron-electron correlations  in valence $\pi$-subsystem of electronic system (but not in $\sigma$-subsystem), since the only  $\pi$-subsystem can be subjected to so strong response (which is mapped by  the strong anisotropy of FMSWR-modes above analysed with the axes coinciding with symmetry axes of the host lattice) to the surrounding lattice presence by NT-formation. The $\sigma$-subsystem of NTs produced is expected to be not so sensitive to  the presence of host diamond lattice. Given prediction is confirmed by formation of own magnetic symmetry with axes, which are not coinciding with symmetry axes of host diamond lattice in NTs, produced by $\left\langle{100}\right\rangle$ ion beam modification. It is explained by direct participation of $\sigma$-subsystem  in magnetic ordering formation, \cite{Yerchuck_Stelmakh_Dovlatova_Yerchak_Alexandrov}.

 It is interesting, that, although NT-formation is starting at  much more lower ion fluences,  the given  strong response to diamond lattice presence by initial NT-formation process is preserved by entire HEIBM of near surface region in diamond single crystal.  
Further, amplitude dependence, Figure 3, is characterised by the clearly pronounced maximum near 35.3 degrees from [100] crystal  direction of diamond lattice, that is,  it coincides also with symmetry direction of diamond lattice. It is seen, that  amplitude dependence  is also strongly anisotropic.  Really, the ratio of maximal amplitude value to its minimal value  is equal to 4.8. It is additional indication, that mechanism of ferromagnetic ordering is connected with the formation of ordered structures just in $\pi$-subsystem of incorporated NTs. Simultaneously, it is the display  of the fact, that anisotropy of FMSWR absorption by magnetically ordered NT-structure and the anisotropy of  NT-structure itself, which is determined by relatively weak g-value anisotropy of main resonance mode, are quantitatively different. 

Two not very pronounced  maxima at  54.7 degrees  and in direction near 17 degrees from [100] crystal are appeared additionally to the maximum near 35.3 degrees direction in the angular dependence of the intensity, Figure 4.  It is clear, that additional maxima are determined by contribution of linewidth anisotropy, compare Figure 4 and Figure 2.

\section{Discussion}

Let us indicate once again, that although it was established, that copper and boron HEIBM with implantation direction along  $\langle{111}\rangle$ crystal axis,  nickel HEIBM with implantation direction along $\langle{110}\rangle$  axis,  and  boron HEIBM of polycrystalline diamond films with implantation direction transversely  to film surface  lead to formation of NTs, incorporated in diamond matrix, which  possess by ferromagnetic ordering, \cite{Erchak}, \cite{Erchak_JETP}, \cite{Ertchak_JAS},
 \cite{Ertchak_Stelmakh},  the nature of given state and mechanism, leading to its formation was not  strictly theoretically described. To solve given task, it seems to be  necessary to know the   nature  of charge and spin carriers and the mechanisms  of carrier transport and interactions of charge and spin carriers  both between themselves and with phonons and  photons. 
There seems to be paramount significant  the same task for nanoelectronics, spintronics  and for the other branches of nanotechnology.  There is existing in the theory of 1D electronic systems in particular in the theory of conducting NTs the following concept, which was starting with the work of Tomonaga in
1950, \cite{Tomonaga}, and  with the work by Luttinger in 1963, \cite{Luttinger},  when it has become clear that the electron-electron
interaction destroys the sharp Fermi surface and leads to a breakdown of the
Landau  Fermi liquid (LFL) theory.  The resulting non-LFL state is commonly called Luttinger
liquid (LL), or sometimes Tomonaga-Luttinger liquid  to describe the universal low-energy properties of
 1D conductors. LL behavior is characterized by
pronounced power-law suppression of the transport current and the density of states,
and by effect of spin-charge separation. The nature of the spin and charge carriers was also proposed to be the following. The chargeless spin 1/2 quasiparticles - spinons and spinless quasiparticles with the charge $\pm$ e  - holons were introduced. It was concluded from the universality of LL description  that the physical properties do
not depend on details of the model, the interaction potential, and so on, but instead they
are only characterized by a few parameters - critical exponents.  Further, the LL concept was
believed to hold for arbitrary statistical properties of the particles, that is,  both for
fermions and bosons. It provided a paradigm for non-Fermi liquid physics. 
When concern the carbon NTs, we have to remark, that there is the following viewpoint.
 The single-wall carbon nanotubes (SWCNTs) are considered in many works to be 1D objects (it is not always correct, especially for standard NTs with diameter in several nanometers and more)  can
be described the only in the frame of  LL concept. Moreover, SWCNTs
are considered even to be the best model system of the LL state demonstration. However, given viewpoint is based the only on  experimental observation of
 power-law behavior  by measuring
the tunneling conductance of SWNTs in dependence on  temperature and voltage andv ballistic nature of transport, which was found by  electron force microscopic measurements   At the same time spin-charge separation by mechanism of spinons and/or holons' formation  has not been observed
so far. Therefore, the existing viewpoint seems to be insufficiently grounded, since power-law behavior of
the tunneling conductance in dependence on  temperature and voltage and ballistic nature of transport phenomena can be explained in the frames of the quite other conceps.

It has to be also remarked, that both the models LL and LFL  are the models of ideal (and even oversimplified) quantum liquids,  since they do not take into account the nonlinearity of the fermion spectrum on the one hand and electron-phonon interactions on the other hand. In fact both the models describe not strongly adequately the real processes, since the changes in charge state of arbitrary atom in 1D chain to be the result of electron-electron interaction are always accompanied by the changes in phonon subsystem (and vice versa). It is consequence of generic coupling between operators of creation and annihilation in electron subsystem and in phonon field. Let us also remember, that key argument for insertion of the notion "Luttinger liquid" is in fact the simplification, determined by linearization of the generic spectrum of particles in neighborhood of Fermi points in k-space. Just given simplification has led to divergencies arising in the perturbation theory in 1D-case and it does not means that 1D Fermi liquid description is incorrect in general case.
Consequently, the description of NTs  the only in the frames of LL concept seems to be also oversimplification. Moreover, it is  showed in \cite{Yerchuck_Dovlatova_Borovik}, that the concept of description of 1D correlated electronic systems in the frame of 1D Fermi liquid (FL) can be renewed. The most interesting result is that,  that the Fermi liquid concept can be applied just to quasi-1D carbon NTs.  
 It was considered in  \cite{Yerchuck_Dovlatova_Borovik} the concept of 1D FL on the example of well known 1D system - \textit{trans}-polyacetylene (t-PL), that is, it is in fact the generalization of well known model developed by Su, Schrieffer, Heeger (SSH-model), \cite{SSH}, \cite{SSH_PRB}, \cite{Heeger_1988}. SSH-model, in distinction from LFL and LL models, takes into account the electron-phonon interaction. The subsequent generalization, for instance, for application  of given model immediately to  quasi-1D carbon zigzag shaped nanotubes can be  easily  obtained by using of hypercomplex number theory like to its application in the works, \cite{Yerchuck_D_Dovlatova_A}, \cite{Dovlatova_Yearchuck}, where hypercomplex number theory was used for the interpretation of quantum optics effects in quasi-1D NTs. 

Let us summarise briefly for the convenience of the readers the  results, presented in  \cite{Yerchuck_Dovlatova_Borovik}.

The Born-Oppenheimer approximation was considered and  starting  Hamiltonian was the following
\begin{equation}
\label{Eq1m}
\hat{\mathcal{H}}(u) = \hat{\mathcal{H}}_{0}(u) + \hat{\mathcal{H}}_{\pi,t}(u) + \hat{\mathcal{H}}_{\pi,u}(u).
\end{equation}
 The first term in 
(\ref{Eq1m}) is  
\begin{equation}
\label{Eq2m}
\begin{split}
\hat{\mathcal{H}}_{0}(u) = \sum_{m}\sum_{s}(\frac{\hat{P}_m^2}{2M^*}\hat{a}^+_{m,s} \hat{a}_{m,s} + K u_m^2 \hat{a}^+_{m,s} \hat{a}_{m,s}).
\end{split}
\end{equation}
It represents   the sum of  operator of kinetic energy of CH-group motion (the first term in (\ref{Eq2m})) and the  operator of the $\sigma$-bonding energy
(the second term). Coefficient $K$ in (\ref{Eq2m}) is effective $\sigma$-bonds spring constant, $M^*$ is total mass  of CH-group, $u_m$ is configuration coordinate for $m$-th CH-group,  which corresponds to translation of $m$-th CH-group along the symmetry axis $z$ of the chain, $m = \overline{1,N}$, $N$ is number of CH-groups in the chain, $\hat{P}_m$ is operator of impulse, conjugated to configuration coordinate $u_m$, $m = \overline{1,N}$, $\hat{a}^+_{m,s}$, $\hat{a}_{m,s}$ are creation and annihilation operators of creation or annihilation of quasiparticle with spin projection $s$ on the $m$-th chain site in  $\sigma$-subsystem of t-PA. 

The second term in 
(\ref{Eq1m}) can be represented in the form of two components and it is

\begin{equation}\begin{split}
\label{Eq3m}
&\hat{\mathcal{H}}_{\pi,t}(u) = \hat{\mathcal{H}}_{\pi,t_0}(u) + \hat{\mathcal{H}}_{\pi,t,\alpha_1}(u) =\\
&\sum_{m}\sum_{s}[(t_0 (\hat{c}^+_{m+1,s} \hat{c}_{m,s} + \hat{c}^+_{m,s} \hat{c}_{m+1,s})]  +\\
& (-1)^m 2 \alpha_1 u )(\hat{c}^+_{m+1,s} \hat{c}_{m,s} + \hat{c}^+_{m,s} \hat{c}_{m+1,s})],
\end{split}
\end{equation}
where $\hat{c}^+_{m,s}$, $\hat{c}_{m,s}$ are creation and annihilation operators of creation or annihilation of quasiparticle with spin projection $s$ on the $m$-th chain site in  $\pi$-subsystem of t-PA. 
It is the resonance interaction (hopping interaction in tight-binding model approximation) of
 quasiparticles in $\pi$-subsystem of t-PA electronic system, which is considered to be Fermi liquid, and
in which the only constant and linear terms in Taylor series expansion of resonance integral about the dimerized state are  taking into account.

The expression for the  operator $\hat{\mathcal{H}}_{\pi,u}(u)$, which describes the part of electron-phonon interaction, determined by interaction  between quasiparticles in Fermi liquid state of $\pi$-subsystem in terms of $\{\hat{c}^{(c)}_{k,s}\}$ and $\{\hat{c}^{(v)}_{k,s}\}$ is the following
\begin{equation}
\begin{split}
\label{Eq11m}
\hat{\mathcal{H}}_{\pi,u,\alpha_2}(u) = \sum_{k}\sum_{k'}\sum_{s}\alpha_2(k, k',s) \hat{c}^{+(c)}_{k',s}\hat{c}^{+(v)}_{k',s}  \hat{c}^{(v)}_{k,s}\hat{c}^{(c)}_{k,s}.
\end{split}
\end{equation}
It has to be indicated, that  the operators $\{\hat{c}^+_{m,s}\}$,   $\{\hat{c}_{m,s}\}$, $m = \overline{1,N}$, were represented in the form
\begin{equation}
\begin{split}
\label{Eq5m}
&\{\hat{c}_{m,s}\} = \{\hat{c}^{(c)}_{m,s}\} + \{\hat{c}^{(v)}_{m,s}\},\\
&\{\hat{c}^+_{m,s}\} = \{\hat{c}^{+(c)}_{m,s}\} + \{\hat{c}^{+(v)}_{m,s}\},
\end{split}
\end{equation} related to $\pi-c$- and $\pi-v$-band correspondingly,
and then $\vec{k}$-space operators were defined
\begin{equation}
\begin{split}
\label{Eq6m}
&\{\hat{c}^{(c)}_{k,s}\} = \{\frac{i}{\sqrt{N}}\sum_{m}\sum_{s}(-1)^{m+1}\exp(-ikma)\hat{c}^{(c)}_{m,s}\},\\
&\{\hat{c}^{(v)}_{k,s}\} = \{\frac{1}{\sqrt{N}}\sum_{m}\sum_{s}\exp(-ikma)\hat{c}^{(v)}_{m,s}\},
\end{split}
\end{equation}
$m = \overline{1,N}$.
The consideration was  restricted  by the taking into account the contribution of the term, corresponding to interaction  between the quasiparticles in different bands, which seems to be the most essential. 
Physically the identification of linear on displacement $u$ part of  resonance interaction (hopping) and the pairwise interaction of quasiparticles in $\pi$-subsystem between themselves with electron-phonon interaction, which was done in cited work, is understandable, if to take into account, that by atomic $CH$ group displacements the phonons  are generated, which in its turn can by release of the place on, for instance, $(CH)_m$ group, to deliver the energy and impulse, which are necessary for transfer of the quasiparticle (electron) from adjacent $(m-1)$- or $(m+1)$-position in chain in the case of  resonance interaction (hopping). For the case the pairwise interaction of quasiparticles, it means, that its linear on displacement $u$ part is realized by means of phonon field,  which transfers the energy and impulse from one quasiparticle to another (which can be not inevitable adjacent). Mathematically it was proved in the following way. The processes of interaction in $c$ ($v$) band can be considered to be independent on each other. It means, that transition probability from the $\langle k_{l,s}|$-state to $\langle k_{j,s}|$-state in $c$-band and from $\langle k'_{l,s}|$-state to  $\langle k'_{j,s}|$-state  in $v$-band, which is proportional to coefficient $\alpha_2(k, k',s)$, can be expressed in the form of product of real parts of corresponding matrix elements, that is in the form
\begin{equation}
\begin{split}
\label{Eq12m}
&\alpha_2(k, k',s) \sim  Re\langle k_{l,s}|\hat{V}^{(c)}|k_{j,s}\rangle Re\langle k'_{l,s}|\hat{V}^{(v)}|k'_{j,s}\rangle = \\
&\sum_{k_{ph}}Re\langle k_{l,s}|\hat{V}^{(c)}|k_{ph}\rangle \langle k_{ph}|k_{j,s}\rangle \times \\
&\sum_{k_{ph}}Re\langle k'_{r,s}|\hat{V}^{(v)}|k_{ph}\rangle\langle k_{ph}|k'_{n,s}\rangle,  
\end{split}
\end{equation} 
where $\hat{V}^{(v)}$ = $V_{0(v)}\hat{e}$ ($\hat{e}$ is unit operator) is the first term in Taylor expansion of pairwise interaction of quasiparticles, for instance, with wave vectors  $k'_{r}$,  $k'_{n}$ and spin projection $s$ in $v$-band, that is, in ground state, $\hat{V}^{(c)}$ = $V_{1(c)} u \hat{e}$ is the second term in Taylor expansion of pairwise interaction in excited state (in c-band), that is, it is   product of configuration coordinate $u$ and coordinate derivative at $u = 0$  of operator of pairwise interaction of quasiparticles with wave vectors $k_{l}$, $k_{j}$ and spin projection $s$   in $c$-band, $k_{ph}$ is phonon wave vector, and the summation is realized over all the phonon spectrum. At that, since the linear density of pairwise interaction is independent on $k$, which is the consequence of translation invariance of the chain, $V_{0(v)}$, $V_{1(c)}$ are constants. Therefore, the  pairwise interaction is considered to be accompanying by process of phonon generation, when electronic quasiparticles are already in excited state, that is,   in $c$-band (retardation effect of phonon subsystem is taken into account). Then it will be  $\hat{V}^{(c)} = {V}_{0(c)}u \hat{e}$, $\hat{V}^{(v)} = V_{0(v)} \hat{e}$.  A number of variants are possible along with process of phonon generation, corresponding to states of electronic quasiparticles in $c$-band above described.  The result will mathematically be quite similar, if to change the energetic place of excitation, that is, if to interchange the role of $c$ and $v$ bands for given process. There seem to be possible the realization of both the stages (that is phonon generation and absorption) for electronic quasiparticles in single $c$ or $v$ band states and simultaneous realisation both the stages in both the bands. Mathematical description will be for all possible variants  similar and for distinctness, it was considered only the first variant. For the simplicity, the processes were considered, in which the spin projection is keeping to be the same. Since in $z$-direction the impulse distribution is quasi-continuous (the chain has  the macroscopic length $L = N a$) the standard way $\sum_{k_{ph}}\rightarrow \frac{L}{2\pi}\int_{k_{ph}}$ has been used. Further, phonon states were described by wave functions $\langle k_{ph}| = v_0 exp(ik_{ph} z)$, where $z \in [0,L]$, $k_{ph} \in [-\frac{\pi}{2a}$, $\frac{\pi}{2a}$], $v_0$ is constant. Then, it was obtained from (\ref{Eq12m})  the expression
\begin{equation}
\begin{split}
\label{Eq13m}
&\alpha_2(k, k',s) = b |v_{0v}|^2 |v_{0c}|^2  V_{0(c)} u V_{0(v)} |\phi_{0cs}|^2 |\phi_{0vs}|^2 \times\\ 
&\frac{N}{2\pi(q_l - q_j)(q_r - q_n)} Re\{\exp[{i(k_l m_l - k_j m_j)a}] \exp{ika}\} \times \\
&Re\{\exp[{i(k'_r m_r - k'_n m_n)a}] \exp {ik'a}\},
\end{split}
\end{equation}
where $|\phi_{0cs}|^2$, $|\phi_{0vs}|^2$ are squares of the modules of the wave functions $|k_{j,s}\rangle$ and $|k'_{j,s}\rangle$ respectively, $k = k_{ph}(q_l - q_j)$, $k' = k'_{ph}(q_r - q_n)$ $q_l, q_j, q_r, q_n \in N$ with additional conditions $(q_l - q_j)a \leq L$, $(q_r - q_n)a  \leq L$, $b$ - is aspect ratio, which in principle can be determined by comparison with experiment. Here
the values $(q_l - q_j)$, $(q_r - q_n)$ determine the steps  in pairwise interaction with phonon participation and they are considered to be fixed.  The processes, for which $k = k'$, were considered. Consequently, $(q_r - q_n)$ =  $(q_l - q_j)$. Then the  Hamiltonian $\hat{\mathcal{H}}_{\pi,u,\alpha_2}(u)$ was represented in the form
\begin{equation}
\begin{split}
\label{Eq16m}
&\hat{\mathcal{H}}_{\pi,u,\alpha_2}(u) = \\ 
&\sum_{k}\sum_{k'}\sum_{s}4 \alpha_2 u \sin ka  \sin k'a \hat{c}^{+(c)}_{k',s}\hat{c}^{+(v)}_{k',s}  \hat{c}^{(v)}_{k,s}\hat{c}^{(c)}_{k,s},
\end{split}
\end{equation}
where $ 4 \alpha_2(s)$ is
\begin{equation}
\begin{split}
\label{Eq15m}
&b |v_{0v}|^2 |v_{0c}|^2  V_{0(c)}  V_{0(v)} |\phi_{0cs}|^2 |\phi_{0vs}|^2 \times \\
&\frac{N}{2\pi[(q_l - q_j)]^2} = 4 \alpha_2(s)
\end{split}
\end{equation}

Let us remark, that the Hamiltonian $\hat{\mathcal{H}}_{\pi,u,\alpha_2}(u)$ describes the attraction between the electrons, it can lead to formation of Cooper pairs in a $\pi$-subsystem  and to superconductivity. 

Two values for the energy of quasiparticles, indicating on the possibility of formation of the quasiparticles of two kinds both in  $c$-band and $v$-band  have been obtained.    They are the following
\begin{equation}
\begin{split}
\label{Eq50m}
&E_k^{(c)}(u) =   \frac{\mathcal{Q}^2 \Delta^2_k  - \epsilon^2_k}{\sqrt{\epsilon^2_k + \mathcal{Q}^2  \Delta^2_k}},\\ 
&E_k^{(v)}(u) =   \frac{\epsilon^2_k - \mathcal{Q}^2 \Delta^2_k}{\sqrt{\epsilon^2_k + \mathcal{Q}^2  \Delta^2_k}}
\end{split}
\end{equation}
 and 
\begin{equation}
\begin{split}
\label{Eq51m}
&E_k^{(c)}(u) =   \sqrt{\epsilon^2_k + \mathcal{Q}^2  \Delta^2_k},\\ 
&E_k^{(v)}(u) =   - \sqrt{\epsilon^2_k + \mathcal{Q}^2  \Delta^2_k}
\end{split}
\end{equation}
 The  factor $\mathcal{Q}$ is determined by relation
\begin{equation}
\begin{split}
\label{Eq36m}
[1 + \frac{\alpha_2}{2\alpha_1} \sum_{k}\sum_{s}\frac{\mathcal{Q} \Delta_k \sin ka }{\sqrt{\epsilon^2_k + \mathcal{Q}^2 \Delta^2_k}} ({n}^{(c)}_{k,s} - {n}^{(v)}_{k,s})] = \mathcal{Q},
\end{split}
\end{equation}
where   ${n}^{(c)}_{k,s}$ is eigenvalue of density  operator of quasiparticles' number in $c$-band,  ${n}^{(v)}_{k,s}$ is eigenvalue of density operator of quasiparticles' number in $v$-band.
The quasiparticles with the energy, determined by (\ref{Eq51m})  at  $\mathcal{Q} = 1$ are the same quasiparticles, that were obtained in known SSH-model. 

Subsequent analysis  has showed,  that   SSH-like solution (\ref{Eq51m}) is inapplicable for the description of standard processes, passing near equilibrium state by any parameters.  The quasiparticles, described by   SSH-like solution, can be created the only in strongly nonequilibrium state with inverse                                                                                                   
population of the levels in $c$- and $v$-bands.                                                                                   
 At the same time  the solution, the energy of quasiparticles for which is determined by (\ref{Eq50m})
 can be realised  both in near equilibrium and in strongly non-equilibrium states of the $\pi$-subsystem of t-PA, which is considered to be quantum Fermi liquid.

The continuum limit for the ground state energy of the $t$-PA chain with SSH-like quasiparticles will coincide with known solution, \cite{SSH_PRB}, \cite{Heeger_1988}, if to  replace $\Delta_k \mathcal{Q} \rightarrow \Delta_k$.   The calculation of   the ground state energy $E^{[u]}_0(u)$ of the $t$-PA chain with  quasiparticles' branch, which is stable near equilibrium by taking into account, that in ground state ${n}^c_{k,s} = 0$, ${n}^v_{k,s} = 1$, in the continuum limit gives
\begin{equation}
\label{Eq57m}
E^{[u]}_0(u) = - \frac{2N a}{\pi}\int\limits_0^{\frac{\pi}{2a}} \frac{(\mathcal{Q}\Delta_k)^2 - 
\epsilon^2_k}{\sqrt{\mathcal{Q}\Delta_k)^2 + 
\epsilon^2_k}}dk + 2NKu^2.
\end{equation}
Then, calculation of  the integral results in the expression
\begin{equation}
\label{Eq58m}
\begin{split}
&E^{[u]}_0(u) =  \frac{4Nt_0}{\pi}\{F(\frac{\pi}{2}, 1 - z^2) + \\
&\frac{1 + z^2}{1 - z^2}[E(\frac{\pi}{2}, 1 - z^2) - F(\frac{\pi}{2}, 1 - z^2)]\} + 2NKu^2, 
\end{split}
\end{equation}
where
$z^2 = \frac{2\mathcal{Q}\alpha_1 u}{t_0}$, $F(\frac{\pi}{2}, 1 - z^2)$ is the complete elliptic integral of the first kind  and $E(\frac{\pi}{2}, 1 - z^2)$ is the complete elliptic integral of the second kind.
Approximation of ground state energy at $z \ll 1$  for the solution, which is stable near equilibrium position, gives
\begin{equation}
\label{Eq59m}
\begin{split}
&E^{[u]}_0(u) = N \{\frac{4t_0}{\pi} - \frac{6}{\pi}\ln\frac{2t_0}{\mathcal{Q}\alpha_1 u} \frac{4 (\mathcal{Q}\alpha_1)^2 u^2}{t_0} + \\
&\frac{28 (\mathcal{Q}\alpha_1)^2 u^2}{\pi t_0} + ...\} + 2NKu^2.
\end{split}
\end{equation}
It is seen from (\ref{Eq59m}), that the energy of quasiparticles, described by  given solution,  has the form of Coleman-Weinberg potential with two minima at the values of dimerization coordinate $u_0$ and $-u_0$ like to the energy of quasiparticles, described by   SSH-solution. It is understandable, that subsequent considerations, including electrically neutral S = 1/2 soliton and electrically charged spinless soliton formation, that is the appearance of the phenomenon of spin-charge separation,  by FL description of 1D systems  will be coinciding in its  mathematical form with those ones in   SSH-model. 

It is  the main result of the work \cite{Yerchuck_Dovlatova_Borovik}, which means, that the physical properties of  1D systems can be described in the frames of 1D quantum FL including the mechanism of appearance of the most prominent feature of 1D systems - the phenomenon of spin-charge separation.

Since the model proposed takes into consideration the electron-electron correlations in explicit form, it seems to be ground for its application to electronic systems, in which electron-electron correlations are  rather strong. In particular, it can be used for analysis of the physical properties including ESR and FMSWR spectra analysis in quasi-1D carbon NTs.  It can be done by above indicated manner, that is by using of hypercomplex number theory analogously to theoretical analysis of quantum optics effects in \cite{Dovlatova_Yearchuck} and  analysis in \cite{Yerchuck_D_Dovlatova_A} of Raman spectra  in given objects. 
 
 It is substantial, that the mechanism of the phenomenon
of spin-charge separation in 1D FL is topological soliton mechanism, which 
  is quite different from Anderson spinon-holon mechanism.

The results obtained show, that 
the shape of $\pi$-solitons (or $\sigma$-solitons) is given by the expression having  the same mathematical form both in  SSH-model and in its FL generalisation. It is
\begin{equation}
\label{Eq16}
|\phi(n)|^2 = \frac{1}{\xi_{\pi(\sigma)}} sech^{2}[\frac{(n-n_0)a}{\xi_{\pi(\sigma)}} - v_{\pi(\sigma)} t] \cos^2 \frac{n \pi}{2},
\end{equation} 
where $n, n_0$ are variable and fixed numbers of $CH$-unit in $CH$-chain, $a$ is $C-C$ interatomic spacing projection on chain direction, $v_{\pi(\sigma)}$ is $\pi$ ($\sigma$)-soliton velocity, $t$ is time, $\xi_{\pi(\sigma)}$ is $\pi$($\sigma$) coherence length. It has been shown in \cite{Yerchuck_Dovlatova_Borovik}, that $\pi$-solitons ($\sigma$-solitons) differ in fact  by numerical value of coherence length in  SSH-model and in its FL description. 

\begin{figure}[t]
\includegraphics[width=0.47\textwidth]{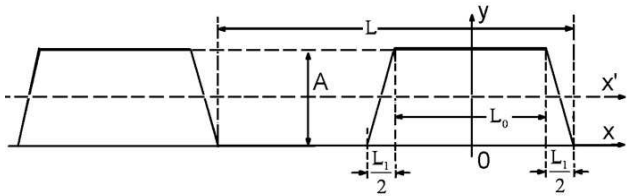}
\caption[Approximation of soliton lattice spin (charge) density wave profile along the chain direction]{\label{Figure15} Approximation of soliton lattice spin (charge) density wave profile along the chain direction}
\end{figure}

The observation of FMSWR phenomenon  in the NTs, formed by $\left\langle{110}\right\rangle$ high energy nickel ion implantation in natural diamond single crystal (at ion beam dose in  $5\times{10^{14}}$ $cm^{-2}$) unambiguously  means, that  the SSH $\pi$-soliton lattice formation takes place, that is periodical non-harmonic spin-density wave. For the simplicity we can approximate the function of the shape of given density wave (let us designate it with abbreviation   ASLSDW  and ASLCDW (Approximation of Soliton  Lattice Spin (Charge) Density Wave) by trapeziform function, Figure \ref{Figure15}. Real function of the envelope shape of soliton  lattice spin density wave will be between given  trapeziform function and harmonic sinusoidal function for density wave (HSDW - Harmonic Spin Density Wave). We will show, that given approximation is rather good for qualitative and in the principle for quantitative estimation of characteristics of ferromagnetic order formation in carbon NTs. The calculation in \cite{Yearchuck_PL} will be adapted for given case. Actually,
 from 
mathematical viewpoint  the difference between ASLSDW and HSDW states is minimal, 
since ASLSDW can be represented in the form of superposition of HSDWs, where amplitude of 
main harmonic is strongly exceeding the others. It follows from  the expansion of ASLSDW profile in Fourier series, which  is
\begin{widetext}
\begin{equation}
\label{eq3}
f(x') = A(2L_0 + L_1)\sum\limits_{m = 0}^\infty  {\left\{ {\frac{{\left[ {2(2L_0  + L_1 )\left( {\cos \frac{{\pi mL_0 }}{{2L_0  + L_1 }} - \cos \frac{{\pi m(L_0+ L_1)}}{{2L_0+L_1}}} \right) - \pi L_1 m\sin m\pi } \right]}}{{2L_1 \pi ^2 m^2 }} \cdot \cos \frac{{2\pi mx'}}{L}} \right\}}, 
\end{equation}
\begin{equation}
\label{eq4}
f(x) = A(2L_0  + L_1 )\left[ {\frac{1}{4} + (2L_0  + L_1 )\sum\limits_{m = 1}^\infty  {\frac{{\left( {\cos \frac{{\pi mL_0 }}{{2L_0  + L_1 }} - \cos \frac{{\pi m(L_0  + L_1 )}}{{2L_0  + L_1 }}} \right)}}{{2L_1 \pi ^2 m^2 }} \cdot \cos \frac{{2\pi mx}}{L}} } \right].
\end{equation}
\end{widetext}

The sense of the parameters L$_{0}$, L$_{1}$, L is seen from Figure \ref{Figure15}. The 
functions $f(x)$, $f(x')$ denote spin density distributions $S(x)$, $S(x')$ in the 
case of ASLSDW and charge density distributions $\rho (x)$, $\rho (x')$ in the 
case of ASLCDW. Expression (\ref{eq3}) corresponds to position of chain atoms along 
$X'$-direction, Figure \ref{Figure15}. Expression (\ref{eq4}) corresponds to position of chain atoms along 
$X$-direction, Figure \ref{Figure15}. We see, that  by asymmetric deviation of 
spin (or charge) density relatively the chain direction,  the constant 
component is appeared. It will lead in the case of  soliton lattice formation to ferromagnetic 
ordering if density wave distribution is magnetic spin  density wave distribution and to 
ferroelectric ordering by similar  electric dipole density  distribution. It is clear from 
(\ref{eq3}) and (\ref{eq4}), that the amplitude of components with $m$ $> 1$ is inversely 
proportional to $m^{2}$, that is, it diminishes rather quickly when number $m$ 
increases. In the first order approximation, the relations (\ref{eq3}) and (\ref{eq4}) can 
be represented by
\begin{equation}
\label{eq5}
f(x') = A\,(2L_0  + L_1 )\,\alpha \cos \frac{{2\pi x'}}{L} + o\left( {\frac{1}{{m^2 }}} \right),
\end{equation}
\begin{equation}
\label{eq6}
f(x) = A\,(2L_0  + L_1 )\left( {\frac{1}{4}\, + \frac{1}{2}\alpha \cos \frac{{2\pi x}}{L}} \right) + o\left( {\frac{1}{{m^2 }}} \right),
\end{equation}
where $\alpha $ is
\begin{equation}
\label{eq7}
\alpha  = \frac{{\left[ {(2L_0  + L_1 )\left( {\cos \frac{{\pi L_0 }}{{2L_0  + L_1 }} - \cos \frac{{\pi (L_0  + L_1 )}}{{2L_0  + L_1 }}} \right)} \right]}}{{L_1 \pi ^2 }}.
\end{equation}
It is seen from (\ref{Eq16}), that envelope function of single soliton spin density distribution corresponds precisely to the case with $x$ axis position, that is, it is strongly  asymmetric relatively the chain axis. By the formation of the soliton lattice the  periodic envelope function of  soliton spin density distribution in soliton lattice will be determined by overlapping of individual envelope functions and the asymmetry extent can be even more pronounced depending on overlapping extent. 

Let us also  remark, that  given conclusion is preserved, when simultaneously with Peierls metal-insulator transition the Mott-Hubbard metal-insulator transition also takes place It leads to modification of oscillating factor in (\ref{Eq16}). Instead of factor  $\cos^2 \frac{n \pi}{2}$ the factor $(b_1 \cos^2 \frac{n \pi}{2}- b_2 \sin^2 \frac{n \pi}{2})$ is appeared, at that in real case of t-PA the value of $b_1$ is substantially exceeds $b_2$, although $b_2$ is nonzero, \cite{Thomann}. 

The viewpoint above developed agrees well with results of the work \cite{Kahol}, where the real lineshapes of SSH-soliton line in t-PA, registered by  both ESR and electron-nuclear double resonance (ENDOR), were approximated even with rectangular spin-density distribution (that is, by particular case of trapeciform distribution at $L_1 = 0$). It seems to be essential, that real ESR lineshape was fitted by given distribution very well by  $\xi_{\pi} = 24a$ and $b_2/b_1 = 0.3$  and qualitatively the lineshape of ENDOR spectrum, which is substantially more sensitive to choose of shape function,  was also reproduced at the same parameters $\xi_{\pi} = 24a$ and $b_2/b_1 = 0.3$.

 Further, it was found also, that there is numerical coincidence of the characteristics of SSH topological $\pi$-solitons in t-PA and in quasi-1D CZSNTS, formed both by $\langle 111 \rangle$ and $\langle 110 \rangle$ ion implantation. So, the values of g-tensor components in $Cu$-implanted sample are $g_1$ = 2.00255 (it is minimal $g$-value and it is $g_{\mid\mid}$ principal direction), $g_2$ = $g_3$ = $g_\perp$ = 2.00273, the accuracy of relative g-value measurements is $\pm 0.00002$, \cite{Erchak}. It is substantial, that $g$-value of paramagnetic $\pi$-solitons in t-PA, equaled to 2.00263, \cite{Goldberg}, gets in the middle of given rather narrow interval of $g$-value variation of PC in ion produced NTs. Although anisotropy of paramagnetic $\pi$-solitons in $t$-PA,  mapping the distribution of $\pi$-electron density along  the whole individual $t$-PA chain, was not resolved by ESR measurements in \cite{Goldberg} directly, there are indirect evidences on axial symmetry of ESR absorption spectra in $t$-PA too, \cite{Kahol}, \cite{Kuroda}. Consequently, the value 2.00263 is average value and it coincides with accuracy 0.00002 with average value of aforecited principal $g$-tensor values of PC in quasi-1D CZSNTs. Taking into account given high precision coincidence of g-values in t-PA and quasi-1D CZSNTs we can conclude that SSH--soliton density distribution in quasi-1D CZSNTs is also asymmetric along hypercomplex $x$-axis. 

Therefore, the appearance of ferromagnetic ordering in quasi-1D CZSNTs, formed by $\langle 111 \rangle$ and $\langle 110 \rangle$ ion implantation, for which the only  $\pi$-subsystem is responsible  is experimentally confirmed.

Thus we have found the physical origin of the mechanism of the formation of ferromagnetic ordering in carbon NTs. It is determined the only by $\pi$-subsystem of full electronic system, which agrees well with experimental results on FMSWR above summarised and explains qualitatively the sensitivity of the symmetry of FMSWR parameters,  Figures 1 to 4, to symmetry directions of surrounding diamond crystal lattice in NTs, produced by $\langle{110}\rangle$ implantation by relative softness of $\pi$-subsystem in comparison with $\sigma$-subsystem. Given conclusion agrees well with the results of the work \cite{Yerchuck_Stelmakh_Dovlatova_Yerchak_Alexandrov}, where the participation of $\sigma$-subsystem in magnetic ordering formation in 
NTs, produced by $\langle{100}\rangle$ implantation, leads to own magnetic symmetry, the  main symmetry axes for which are not coinciding with symmetry axes of the host diamond  lattice.

\section {Conclusions} 

The physical origin of the mechanism of the formation of ferromagnetic ordering in carbon nanotubes  produced by high energy ion beam modification of diamond single crystals in $\langle{110}\rangle$ and $\langle{111}\rangle$ directions has been established. It is determined by asymmetry of spin density distribution  of Su-Schrieffer-Heeger topological soliton lattice, which is formed in 1D Fermi quantum liquid state of  the only $\pi$-electronic subsystem of given NTs. The conclusions are based on detailed analysis of angular dependencies of the parameters of ferromagnetic spin wave resonance, numerical values and symmetry of g-values  of main FMSWR moxdes, taking into account the theory of 1D Fermi liquid and adaptation of the calculation, which gives the qualitative evaluation  of total magnetization and  indicates on the appearance of ferromagnetic ordering by the formation of $\pi$-electronic SSH-soliton lattice (produced by spin 1/2 SSH solitons in 1D Fermi $\pi$-electronic liquid) with asymmetric spin distribution relatively 1D hypercomplex symmetry axis of NTs studied.

\end{document}